\documentclass[lettersize,journal]{IEEEtran}
\usepackage{amsmath,amsfonts}
\usepackage{algorithmic}
\usepackage{algorithm}
\usepackage{array}
\usepackage[caption=false,font=small,labelfont=rm,textfont=rm]{subfig}
\usepackage{textcomp}
\usepackage{stfloats}
\usepackage{url}
\usepackage{verbatim}
\usepackage{graphicx}
\usepackage{booktabs}
\usepackage{cite}
\usepackage{tabularx}
\usepackage[table,xcdraw]{xcolor}
\usepackage{indentfirst}
\usepackage[table]{xcolor}
\usepackage{color}
\usepackage[table]{xcolor}
\usepackage{colortbl}
\usepackage{multirow}

\begin{document}

\title{A  Revolution of Personalized Healthcare: \\ Enabling Human Digital Twin with Mobile AIGC}

\author{Jiayuan Chen, Changyan Yi,~\IEEEmembership{Member,~IEEE,} Hongyang Du, Dusit Niyato,~\IEEEmembership{Fellow,~IEEE}, Jiawen Kang
\\ Jun Cai,~\IEEEmembership{Senior Member,~IEEE,} and Xuemin (Sherman) Shen,~\IEEEmembership{Fellow,~IEEE}

 \thanks{J. Chen and C. Yi are with the College of Computer Science and Technology, Nanjing University of Aeronautics and Astronautics, China (Email: jiayuan.chen@nuaa.edu.cn, changyan.yi@nuaa.edu.cn).
 	
H. Du and D. Niyato is with the School of Computer Science and Engineering, Nanyang Technological University, Singapore (Email: HONGYANG001@e.ntu.edu.sg, dniyato@ntu.edu.sg).

J. Kang is with the School of Automation, Guangdong University of Technology, China (Email: kavinkang@gdut.edu.cn).

J. Cai is with the Department of Electrical and Computer Engineering, Concordia University, Montreal, Canada (Email: jun.cai@concordia.ca).

X. Shen is with the Department of Electrical and Computer Engineering, University of Waterloo, Canada (Email: sshen@uwaterloo.ca).
}
}

\maketitle

\begin{abstract}
Mobile Artificial Intelligence-Generated Content (AIGC) technology refers to the adoption of AI algorithms deployed at mobile edge networks to automate the information creation process while fulfilling the requirements of end users. Mobile AIGC has recently attracted phenomenal attentions and can be a key enabling technology for an emerging application, called human digital twin (HDT). HDT empowered by the mobile AIGC is expected to revolutionize the personalized healthcare by generating rare disease data, modeling high-fidelity digital twin, building versatile testbeds, and providing 24/7 customized medical services. To promote the development of this new breed of paradigm, in this article, we propose a system architecture of mobile AIGC-driven HDT and highlight the corresponding design requirements and challenges. Moreover, we illustrate two use cases, i.e., mobile AIGC-driven HDT in customized surgery planning and personalized medication. In addition, we conduct an experimental study to prove the effectiveness of the proposed mobile AIGC-driven HDT solution, which shows a particular application in a virtual physical therapy teaching platform. Finally, we conclude this article by briefly discussing several open issues and future directions.
\end{abstract}
\section{Introduction}

\subsection{Mobile AIGC-Driven Human Digital Twin}
\IEEEPARstart {H}uman digital twin (HDT), as a recently defined concept, is expected to characterize the replication of an individual human body in the virtual space, while reflecting its physical status both psychologically and physiologically in real-time \cite{1}. The corresponding virtual entity is called virtual twin (VT), while the physical one is called physical twin (PT). HDT is poised to revolutionize personalized healthcare dramatically through its unparalleled capabilities, such as acting as ultra-realistic, human-like, and versatile testbeds, as well as continuously and pervasively monitoring and predicting medical conditions. However, the successful implementation of HDT in personalized healthcare hinges greatly on the utilization of vast amounts of multi-modal data.

Obviously, acquiring such large amounts of data as the input for HDT can be significantly challenging especially when it comes to the individual-level. Traditional data collection methods \cite{10} are often difficult to be personalized and scaled, particularly within the healthcare realm due to the high privacy of medical records. Fortunately, the cutting-edge technology, namely Artificial Intelligence-Generated Content (AIGC) \cite{4}, which has the abilities of creatively generating, manipulating, and modifying valuable and diverse data by using advanced AI algorithms, can be a promising solution for powering up HDT in personalized healthcare. Considering that the healthcare services typically need to be timely, pervasive and uninterrupted, in this article, we investigate the deployment of AIGC at mobile edge networks for HDT, referred to as mobile AIGC-driven HDT, to provide low-latency and interaction-intensive personalized healthcare services.

\begin{enumerate}
	\item \textbf{Generating Rare Data for HDT Utilizing Mobile AIGC}: Traditional data collection methods in HDT for personalized healthcare applications is commonly time-consuming, costly or invasive, especially when dealing with hard-to-reach populations or rare diseases that may affect a small subset of the population. Mobile AIGC-driven data generation can overcome such a data scarcity issue by generating massive synthetic data based on statistical characteristics and patterns of the collected rare data. For instance, NVIDIA has unveiled a new multi-modal (CT/MR/ultrasound) generative AI recently, called RadImageGAN, for generating medical data with considerably small amounts of real ones\footnote{https://www.nvidia.com/en-us/on-demand/session/gtcspring23-s51264/}.

	\item \textbf{Modeling High-Fidelity HDT Utilizing Mobile AIGC}: Traditional digital modeling technologies may fall short of meeting the stringent requirements of HDT. In particular, data-based modeling technologies, such as point-cloud modeling, require extensive individual-level data to create human digital structures. However, it may lack generality despite the high level of accuracy. For example, by this way, modeling viral proteins for typical viruses with abundant historical data is easily achievable, but modeling rare ones becomes much more challenging. Conversely, model-based technologies, such as computer-aided design (CAD), highly rely on strict assumptions, which may experience low accuracy despite the generality. AIGC can well balance these two metrics due to its strong ability in both data generation and analysis. For instance, BioMap recently launched an AI generated protein platform which can produce customized protein by exploiting high-dimensional data and high-performance biological operation mechanisms\footnote{https://www.biomap.com/en/}. It is worth noting that the implementation of such large-scale AIGC models originally require high-intensive computing capacities and fast-responsive outcome feedbacks. These can hardly be fulfilled by only local devices with limited resources or remote clouds in far distance, and thus necessitating the deployment of mobile AIGC supporting by the collaboration of clouds, edge servers and even the local devices.
	\item \textbf{Mobile AIGC-driven HDT for Personalized Versatile Testbed}: Mobile AIGC-driven HDT is envisioned to serve as a versatile testbed for healthcare applications, such as personalized treatment planning. Specifically, multiple candidate treatments can be tested on HDT to identify the best-performing one before actual treating. These scenarios require massive multi-modal feedbacks, such as haptic and drug reactions, for providing human-like responses. AIGC has the potential to help generate those responses. For instance, HaptX, is currently working towards realizing tactile Internet by using AIGC to generate haptic signals (e.g., haptic feedbacks in physical therapy and virtual surgery)\footnote{https://haptx.com}. Additionally, Evozyne and NVIDIA are collaborating to predict and simulate the interactions of drugs in the human body using AIGC\footnote{https://www.nvidia.com/en-us/on-demand/session/gtcspring23-S51713/}. For enabling doctors, pharmacists and patients to interact with HDT for performing personalized testbed functions at anytime and anywhere, deploying AIGC at mobile edge networks is also necessary for guaranteeing seamless connectivities.
	\item \textbf{Mobile AIGC-driven HDT for 24/7 Customized Healthcare Services}: Mobile AIGC-driven HDTs of doctors deploying at mobile edge networks can be trained and served as personal 24/7 doctors for patients. By leveraging mobile AIGC, HDTs of doctors can be empowered with expert-level medical knowledge, providing tailored healthcare services for anybody on demand. For example, Google's ongoing research on Med-PaLM utilizes expert-level medical large language models to accurately and safely answer medical questions\footnote{https://sites.research.google/med-palm/}. In the coming future, mobile AIGC-driven HDTs are expected to offer immersive healthcare through generating multi-modal information in the form of not only text but also audio, video and haptic. In this regard, patients can obtain customized medical services while having both indoor and outdoor activities.
\end{enumerate}

\begin{figure*}[!t]
	\centering
	\includegraphics[width=0.97\textwidth]{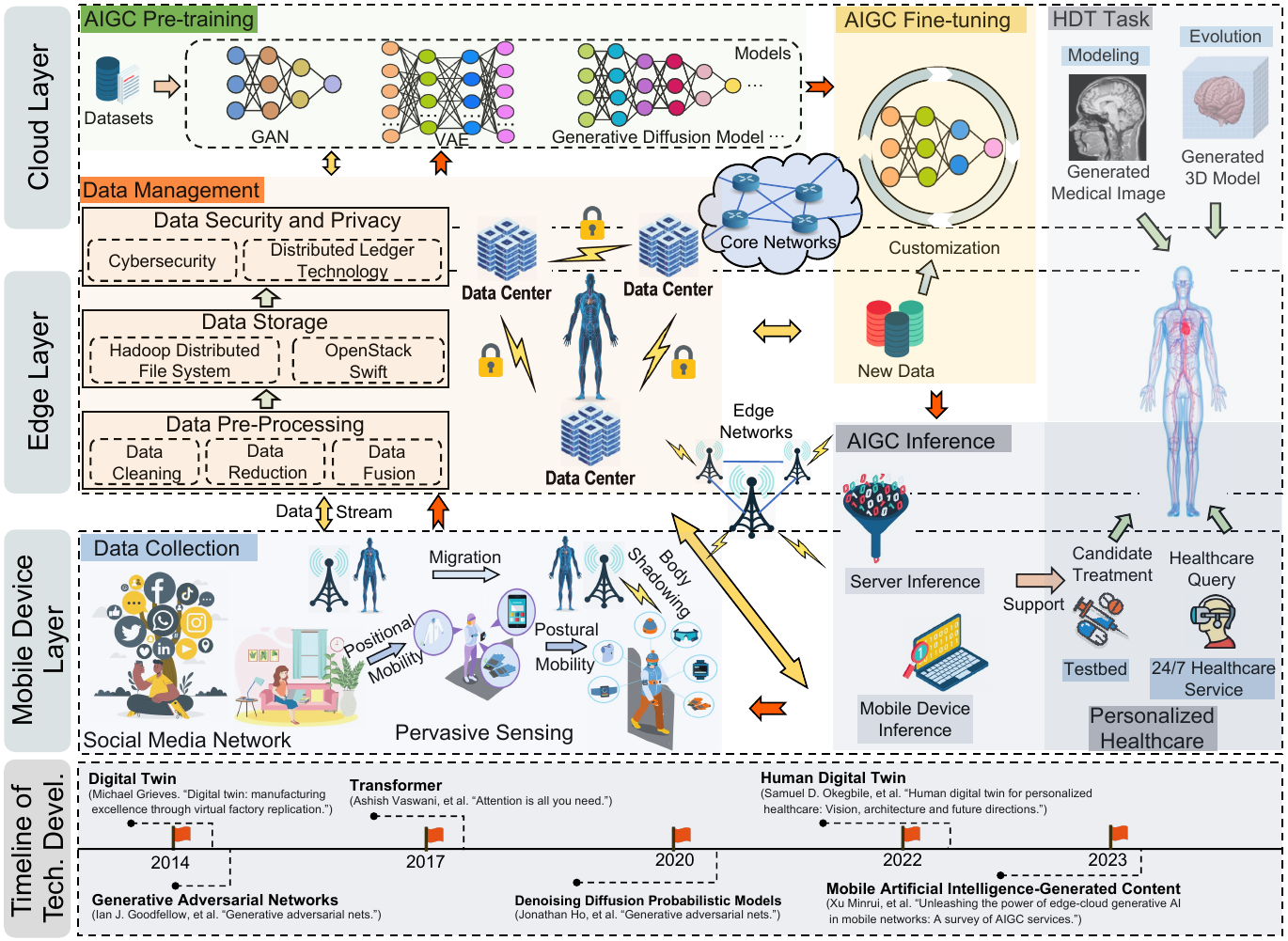} \\
	\caption{The system architecture of mobile AIGC-driven HDT in personalized healthcare,  including the mobile device, edge and cloud layers, attached with the timeline of major technical developments. All these support the mobile AIGC-driven HDT lifecycle, i.e., data collection, management, pre-training, fine-tuning and inference.}\label{Net_Ar}
\end{figure*}

\subsection{Related Work and Contributions}
Both HDT and mobile AIGC have recently attracted a myriad of attentions. For example, Okegbile et al. in \cite{1} defined a general HDT framework, along with key design requirements and corresponding technologies. Chen et al. in \cite{2} clarified and surveyed the enabling technologies of HDT in the perspective of networking. Taylor et al. in \cite{4} provided a general overview of AIGC, discussing its direct applications including image, video and text synthesis. Inspired by this pioneering work, Alamir et al. in \cite{16} delved into studying generative adversarial networks (GANs) in medical image analysis, while Kazerouni et al. in \cite{19} surveyed applications of diffusion models in the medical community. However, none of existing work recognized the potential of AIGC in empowering HDT for the revolution of healthcare.

This motivates us to compose this article that particularly discusses how HDT can be further enabled by mobile AIGC in personalized healthcare. The main contributions of this article are summarized as follows.
\begin{itemize}
	\item We are the first to propose a holistic system architecture of mobile AIGC-driven HDT, along with its lifecycle, including data collection, data management, pre-training, fine-tuning, inference and product management, which can provide insights into the practical implementation.
	\item We rigorously analyze key design requirements and challenges of mobile AIGC-driven HDT in personalized healthcare, which can guide a roadmap for the realization.
	\item We illustrate two use cases of mobile AIGC-driven HDT in personalized healthcare. We also conduct a case study of mobile AIGC-driven HDT in a virtual physical therapy teaching platform as a preliminary exploration.
	\item  We conclude this article by discussing some open issues to inspire future research directions.
\end{itemize}

\section{Framework of Mobile AIGC-Driven HDT in Personalized Healthcare}

\subsection{System Architecture and Key Techniques}

The system architecture of mobile AIGC-driven HDT consists of the mobile device layer, edge layer and cloud layer, as shown in Fig. \ref{Net_Ar}. Besides, the timeline of major technical developments for supporting this architecture is also attached in this figure. The lifecycle of mobile AIGC-driven HDT, including data collection, management, pre-training, fine-tuning and inference, is circulated among the core and edge networks, and is elaborated as follows.
\subsubsection{Data Collection}
Data collection is a critical initial step in mobile AIGC-driven HDT, which is basically carried out by the mobile device layer. The collected data used to train mobile AIGC models directly influence the patterns and relationships that models can learn, and therefore impact the performance in particular applications. For example, electronic health records (EHR) are important training datasets for mobile AIGC-driven HDT \cite{4}. However, EHR datasets are generally confined to specific hospital systems and cannot capture all different kinds of samples, hence limiting the availability of sufficient rare disease patients in the dataset. To this end, administrative claims datasets with wider patient samples can be an alternative to train a transformer-based model for rare disease diagnosis \cite{23}. Except collecting data from these public datasets, the personalization of mobile AIGC-driven HDTs requires data collected from mobile PT themselves. It is generally realized by pervasive sensing using smart biomedical devices, including wearable and implantable devices, for gathering dynamic electroencephalographic (EEG), electrocardiograms (ECG), and biomarkers signals, among others \cite{2}. Additionally, social networks can be utilized as a psychology-related data source for aiding the establishment of mobile AIGC-driven HDT.
\subsubsection{Data Management}
Data from both physical and virtual spaces, including collected data, data generated by AIGC models, simulation data, historical data, etc., are large-scaled and complex. Thus, efficient and reliable data management is indispensable. First, data from each segment should go through pre-processing phase. For instance, diffusion models and GAN have been applied to pre-process medical data, including image-to-image translation, reconstruction, registration, classification, segmentation and denoising \cite{19, 16}. Then, these pre-processed data need to be stored robustly using big data storage frameworks for AIGC model training, data sharing, etc. Moreover, the data in mobile AIGC-driven HDT are highly sensitive. Any leakage of these data may result in serious ethical and moral concerns.

\subsubsection{AIGC Pre-training}
The collected data are used to train AIGC models after being processed by data management. AIGC pre-training is typically done by central servers with powerful computing power at the cloud layer. During training, the generative models, such as GANs \cite{16}, variational autoencoder (VAE) \cite{22}, and generative diffusion models \cite{19}, can automatically learn features of the collected data. Note that, the selection of a specific generative AI technology depends on various factors, including the requirements of tasks, the amount of data, desired output and available resources. For instance, generative diffusion models can learn the inverse diffusion process to produce new  medical images from the artificially added noises, and because of the strong mode coverage and high quality of the generated content, they become popular in medical image synthesis \cite{19}.

\subsubsection{AIGC Fine-tuning}
AIGC fine-tuning is the process of adjusting a pre-trained AIGC model to new tasks, e.g., generating synthetic magnetic resonance imaging (MRI) for rare diseases, by utilizing a modest amount of data. This approach can be applied to enhance the model's performance on given tasks, such as HDT modeling and personalized healthcare services, by slightly modifying AI model's parameters to suit the new data. In mobile networks, AIGC fine-tuning can be performed at the  cloud or edge layer, using the datasets collected from mobile devices and meticulously processed by the data management.
\subsubsection{AIGC Inference}
Based on the trained AIGC models, inference is mainly to generate the desired content according to the input. Since inference commonly requires much less computing power compared to the pre-training, it is expected to install AIGC inference on edge servers, or mobile devices directly to provide low-latency and private AIGC services.  The contents generated by AIGC models can be used to support HDT tasks in personalized healthcare. For example, generating medical images and 3D models can facilitate HDT modeling and evolution, while the generated responses can enable testbed functions and ubiquitous medical services \cite{22}.

\subsection{Key Design Requirements and Challenges}
The key design requirements and major challenges of implementing mobile AIGC-driven HDT in personalized healthcare are discussed as follows.
\subsubsection{Personalized and Adaptive Self-evolution of Mobile AIGC-driven HDT}
To keep synchronized between each PT-VT pair, mobile AIGC-driven HDT needs to be real-time self-evolved once the status of the PT is changed according to the collected data. Such a process is intuitively personalized because the characteristics of each PT-VT pair are totally different. Moreover, it is also adaptive because this process highly depends on the available resources, including computational and communication resources, data accessibility, etc. In the following, we outline several critical points related to this.

\textbf{Adaptive Computing and Networking Resource Orchestration for Heterogeneous Tasks}:
Heterogeneous AIGC-driven HDT tasks in mobile networks may have different quality of service (QoS) requirements, such as accuracy, inference latency and model size, thereby demanding different amounts of computing and networking resources.
Additionally, as the natural progression of human lifecycle continues, AIGC models will become increasingly sophisticated, leading to an exponential increase in the number of parameters and computational complexity (e.g., RadImageGan is usually trained on 8 NVIDIA A100 GPUs for more than 768800 hours).
Moreover, the high mobility of AIGC-driven HDT users results in varying computing and networking resource demands across locations. All these pose great challenges for unified resource management, including measuring, perceiving, and adaptively allocating computing and networking resources. One potential solution may be utilizing network slicing-based resource allocation for fulfilling various mobile AIGC-driven HDT tasks with multi-objectives.

\textbf{Timely Data Collection, Generation and Integration for Dynamic Fine-Tuning}:
The dynamic fine-tuning of mobile AIGC-driven HDT for personalized self-evolution requires the timely data collection, generation and integration. On one hand, the variation of human body will generate new data, and these data should be transmitted to the edge servers and integrated timely for AIGC models' fine-tuning, such that AIGC models can be updated and evolved synchronously \cite{2}. On the other hand, the fine-tuned AIGC models in HDT should generate latest synthesis rare disease-related data for amplifying the training datasets, and further promoting the fine-tuning of the disease-related models. Such dynamic fine-tuning may require the support of 6G communications and Time-Sensitive Network (TSN).

\textbf{Pervasive Connectivity for Ubiquitous Mobility}:
The highly complex mobility patterns of end users in mobile AIGC-driven HDT pose several unique requirements and challenges for adaptive self-evolution of the system. Specifically, these patterns can be categorized into human positional and postural mobility. Positional mobility, like a person moving from indoor to outdoor, may cause radio frequency (RF) propagation characteristics to change and even the service migrations, and thereby, leading to signal fading among other extra network overheads. Additionally, different from other mobile terminals, the postural mobility of a person, like lying, sitting, walking, may cause signal strength to fluctuate due to the influence of human bodies on the path loss, known as body shadowing, leading to service interruptions or packet loss. These factors can impact the reliability, stability, security, and cost of model self-evolution for mobile AIGC-driven HDT. To address these issues, future networking techniques are expected to be designed with the goal of providing pervasive connectivity, so that enormous data traffics between virtual and physical spaces can be unimpededly circulated.

\subsubsection{Customized and Multi-modal Intelligent-interaction of Mobile AIGC-driven HDT}
Mobile AIGC-driven HDT can disruptively act as personalized versatile testbeds and personal 24/7 doctors. In this regard, VTs can generate customized feedbacks adhering to their corresponding PTs for users (e.g., doctors, pharmacists, and patients). In addition, users can gain immersive experience by utilizing multi-modal interactions, such as audio, video and touch, while engaging with the VTs in the virtual scenes (e.g., surgery). In the following, we outline several critical points related to this.

\textbf{Multi-modal Data Processing and Transmission in Immersive Interaction}: With the versatile functions of mobile AIGC-driven HDT, the data collected, processed, and generated in such a system are typically multi-modal, including auditory, visual and haptic signals, such as in the application of immersive rehabilitation. The processing and transmission of such multi-modal data introduce several open issues. Firstly, multi-modal encoding schemes must be properly defined. Although auditory and visual data encoding has been studied extensively, haptic encoding is still very challenging \cite{8}. Secondly, the transmission of auditory, visual, and haptic data during feedback results in multiple data streams that should be synchronized to avoid motion sickness. For example, the time interval between perceived visual and tactile movement should not exceed 1 ms \cite{8}. Besides, the requirements of data transmissions in mobile AIGC-driven HDT depend on different use cases. For instance, in surgery simulation, auditory, visual and haptic data need to be delivered within an end-to-end latency $<$ 1 ms and with an extremely high reliability $> 99.999999\%$, while immersive virtual outpatient service requires the video signals to be transmitted with an air latency $0.5-2$ ms, and a reliability $> 99.999\%$, but the audio and haptic signals only need to be transmitted with a reliability $> 99.9\%$ \cite{8}.

\textbf{Data Privacy, Security and Integrity with Ethics and Morality}: Massive data will be transmitted and exchanged between PTs and VTs in mobile AIGC-driven HDT. Any leakage of these data may result in serious ethical and moral concerns. Note that, for training large-scale AIGC models, decentralized computing techniques are commonly employed to distribute the data across multiple computing nodes to relive the operation pressure \cite{4}. However, during this process, if appropriate data privacy, security, integrity, and availability schemes are not taken, attackers may obtain data by eavesdropping the network traffics and maliciously attacking computing nodes. Therefore, a series of privacy, security and integrity techniques, such as differential privacy, secure multi-party computation, and homomorphic encryption, should be designed and implemented for mobile AIGC-driven HDT in personalized healthcare applications.

\textbf{Integration of Subjective and Objective Evaluations in Personalized Healthcare Application}: Different from the conventional network systems, mobile AIGC-driven HDT requires both subjective and objective metrics to evaluate the performance, including not only the AIGC models' inference and feedback latency, reliability of communication and computation, but also the experience and perception of users in interactions. The subjective measures of contents (e.g., healthcare dialogues, images and videos) generated by AIGC models are important metrics to aid models' performance promotion, especially for personalized AIGC models. For example, although tools like ChatGPT is capable of generating content that usually appears or sounds reasonable, they are often unreliable in terms of factuality. Thus, subjective knowledge and expertise from healthcare professionals should be introduced to reassess the results.
However, it is difficult to design integrated evaluation methods that can jointly capture both subjective and objective aspects, as it may involve interdisciplinary knowledge and complex models.

\section{Application of Mobile AIGC-driven HDT in Personalized Healthcare}
\begin{figure}[!t]
	\centering
	\includegraphics[width=0.9999\columnwidth]{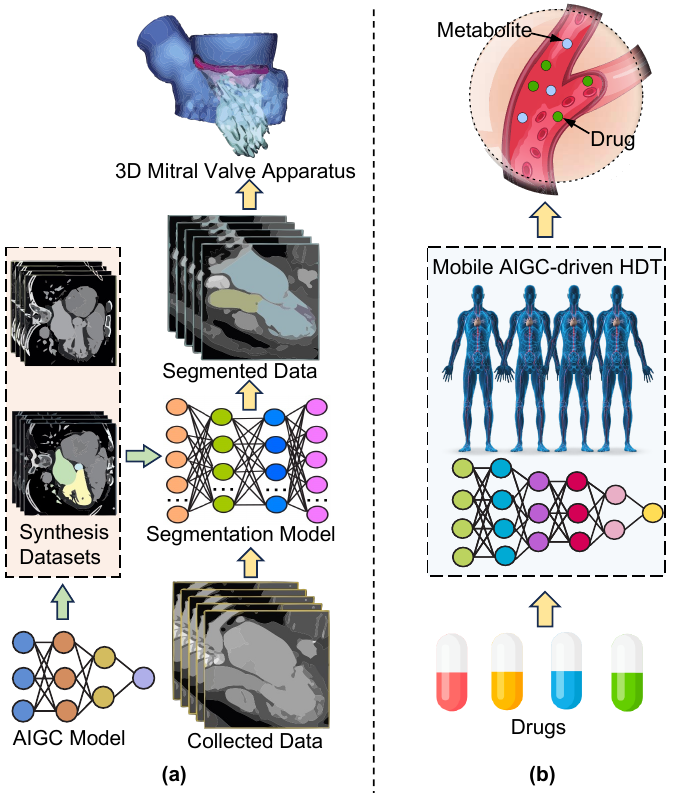} \\

	 \caption[Caption for list of figures]{Use cases of mobile AIGC-driven HDT in personalized healthcare.}
	\label{Use}
\end{figure}

\subsection{Mobile AIGC-driven HDT in Customized Surgery Planning}
One of the potential applications of mobile AIGC-driven HDT in personalized healthcare is customized surgery planning. For example, as shown in Fig. \ref{Use} (a), through the use of high-fidelity 3D models of the heart (e.g., 3D mitral valve apparatus) in VT, physicians can comprehend the morphology, distribution, and vascular pathways of lesions, thereby empowering them to develop more individualized and precise treatment plans. The accuracy of VT modeling is pivotal in ensuring the effectiveness of customized surgery planning, which, in turn, depends heavily on the quality of segmented medical imaging data.
However, due to high privacy of medical data and difficulty of data annotation, existing labeled datasets typically have limited scale, thereby leading to the challenge of training segmentation models with high accuracy and strong generalization ability \cite{10}.

Fortunately, synthesis datasets generated by AIGC models for expanding the training datasets has proven to be a promising solution\footnote{https://subtlemedical.com/subtlesynth/}. For instance, Subramaniam et al. in \cite{18} applied WGAN to generate 3D Time-of-Flight Magnetic Resonance Angiography (TOF-MRA) patches with their corresponding brain blood vessel segmentation labels, eliminating the need for manual annotation and the possibility of privacy leakage. Except for data synthesis, mobile AIGC-driven HDT can aid customized surgery planning through image reconstruction and data augmentation. For example, Song et al. in \cite{24} leveraged the score-based model to reconstruct images from sparse-view lung CT and undersampled brain tumor MRI for producing high-quality medical images in surgery planning. Uemura et al. in \cite{26} developed a flow-based generative model for performing 3D data augmentation of colorectal polyps for effective training of deep learning in computer-aided detection (CADe) for CT colonography. Both these approaches substantially improved the performance of the customized surgery. Furthermore, due to the demand of strong interactions in such surgery planning, it is expected to deploy the inference segment of mobile AIGC-driven HDT at the edge of networks for providing real-time immersive experiences.
	\begin{table*}[htbp]
	\centering
	\caption{Application of Mobile AIGC-driven HDT in Personalized Healthcare}
	\label{usecase}
	
	\includegraphics[width=0.9\textwidth]{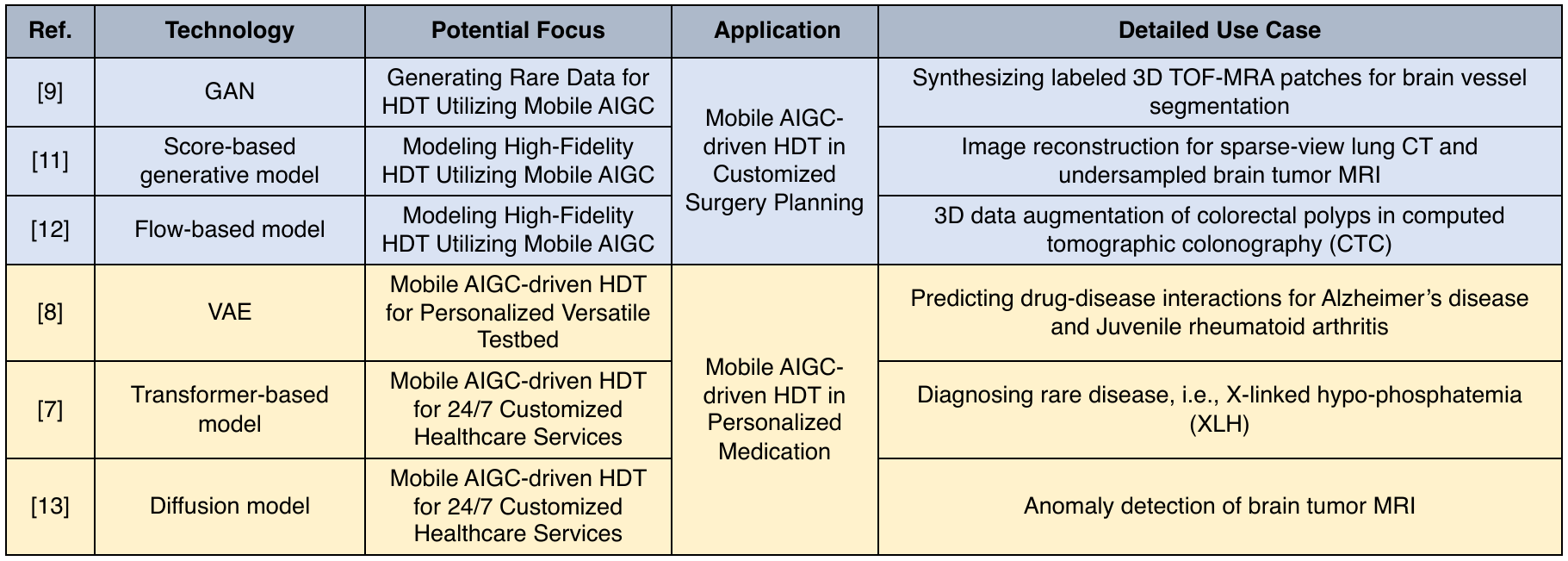}
\end{table*}

\subsection{Mobile AIGC-driven HDT in Personalized Medication}
With the excellent simulation capability and creativity, mobile AIGC-driven HDT can serve as a hyper-realistic and hyper-intelligent testbed for optimizing physical fitness. As shown in Fig. \ref{Use} (b), doctors can prescribe medication by virtually testing various possible prescriptions on the patient's VT. The AIGC models in HDT will then generate simulations of the metabolic processes, efficacy, and toxic side effects of drugs at different doses or drugs in the patient's body based on its unique conditions\footnote{https://www.simulations-plus.com/software/admetpredictor/ai-driven-drug-design-aidd/}. By analyzing the simulation results from mobile AIGC-driven HDT, doctors can further optimize patient treatment plans, taking into account factors such as slower metabolism or excessive reactions to certain drugs due to genetic factors.
	
In this regard, Jarada et al. in \cite{22} proposed a VAE-based approach to predict drug-disease interaction of drug candidates for potentially treating Alzheimer's disease and Juvenile rheumatoid arthritis. Apart from testing candidate prescriptions, mobile AIGC-driven HDT can also aid personalized medication through diagnosing rare diseases and anomaly detection, which can largely improve the efficiency and effectiveness of personalized medication. For example, Prakash et al. in \cite{23} devised a transformer-based approach, called RareBERT, aiming to diagnose rare disease, e.g., X-linked hypo-phosphatemia (XLH). Wyatt et al. in \cite{25} developed a novel diffusion model-based anomaly detection strategy for brain tumor MRI. Based on these, patients themselves may access mobile AIGC-driven HDTs deployed at the edge to obtain timely and customized healthcare services 24/7, no matter where they are and what they are doing. All these ultimately facilitating the personalized medication.

For clarity, Table \ref{usecase} provides a brief survey summarizing different applications of mobile AIGC-driven HDT in personalized healthcare realm.
\begin{figure*}[!t]
	\centering
	\includegraphics[width=0.95\textwidth]{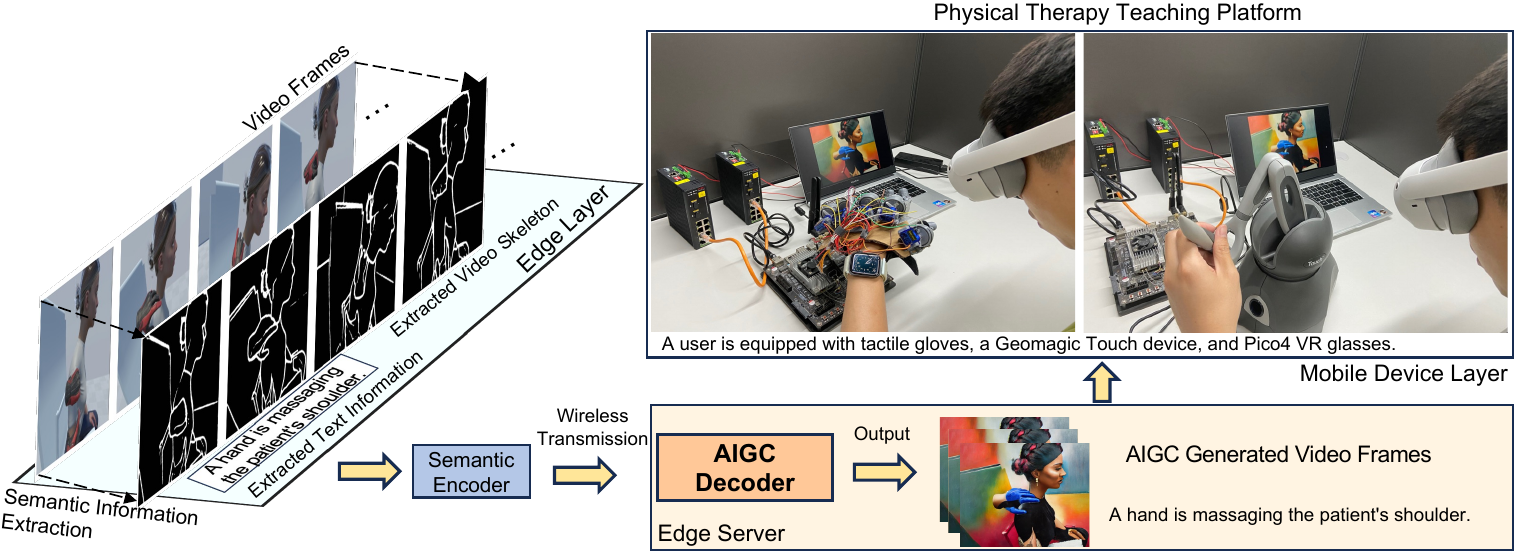} \\
	\caption{The constructed virtual physical therapy teaching platform by applying mobile AIGC-driven HDT.}\label{Cas}
\end{figure*}

\section{A Case Study of Mobile AIGC-driven HDT in Personalized Healthcare}

\subsection{Experimental Setup and System Design}
As shown in Fig. \ref{Cas}, we build a virtual physical therapy teaching platform by applying mobile AIGC-driven HDT. Specifically, the platform is deployed on an edge server, where numerous users, including trainers and trainees, can access the server for participating the virtual class. Users (i.e., PTs) engage the virtual class through their VTs. As the virtual scene shown in Fig. \ref{Cas}, the hand represents a hand of a trainer or trainee VT, and the woman represents a patient VT. The trainer or trainee PT controls his/her VT to massage the shoulder of the patient VT, and then the platform provides all PTs with immersive experience by feeding back haptic signals and VR video streams, among other feedback signals. The haptic signals are transmitted by tactile Internet \cite{8}, to users' tactile devices (e.g., Geomagic Touch or tactile gloves). Furthermore, in this highly interactive scenario where low-latency transmission is required, the VR video streams are typically large in size. To address this challenge, we utilize semantic communication to deliver VR video streams with the goal of reducing data size, network overhead and communication latency. Particularly, as shown in Fig. \ref{Cas}, the VR video streams will go through semantic information extraction in the edge layer, including video skeleton and text extractions (e.g., a short video that a pair of hands is massaging a woman), and then those extracted information are encoded by a video semantic encoder before being delivered to users. To simultaneously cope with the issue that VR videos may be corrupted due to healthcare data collection errors and transmission packet losses, at the user side' edge server, we deploy an AIGC model using the received extracted information and video skeleton to generate a video stream that closely resembles the actual one. This paradigm demonstrates that the powerful generative ability of AIGC cannot only enable the provision of personalized healthcare services \cite{18, 24, 26, 25}, but also enhance the communication efficiency in the PT-VT interactions of HDT.

Although such mobile AIGC-driven paradigm is well-functioned and cost-efficient, it may suffer from a degradation of users' quality of experience (QoE), which is defined as the linear combination of the bitrate of VR video and the similarity between the generated  and  actual ones. To balance the trade-off between QoE and network overhead, we further formulate an optimization problem on top of the constructed testbed (i.e., virtual physical therapy teaching platform), taking into account the constraints of bandwidth resource, computation resource and users' QoE thresholds, in optimizing the resolution ratio and the diffusion step, with the aim of maximizing the total users' QoE. Here, the diffusion step is a key parameter in generative diffusion models, which depicts the number of step of removing Gaussian noise by running a neural network (e.g., U-Net) \cite{19}.
Note that the resolution ratio and diffusion step are proportional to the bitrate and similarity, respectively.

\subsection{Solution and Result Analysis}

A conditional diffusion model-based approach (CODI), with the basic idea following \cite{27},  is particularly proposed to address the optimization problem formulated above. In general, for any given environment (i.e., the status of bandwidth resource, computation resource and users' QoE thresholds), decisions (i.e., each user's resolution ratio and diffusion step) are first randomly generated according to the standard Gaussian distribution. After the multi-step iterative denoising through the proposed CODI approach, the output is the optimal decision that maximizes the total users' QoE. To be more specific, we employ the actor-critic architecture-based deep reinforcement learning (DRL) paradigm to train the conditional diffusion model, where the conditional diffusion model network acts as the actor to map a given environment to the optimal decisions, and the critic network is trained by the double Q-learning technique to evaluate the value of decisions made by the actor.

We compare the performance of the proposed CODI approach with two conventional DRL algorithms, i,e., SAC and PPO \cite{21}, in the constructed testbed. Fig. 4 (a) illustrates the reward (i.e., reflected by the total users' QoE) obtained by three different algorithms during the training stage. It is shown that PPO requires more training steps to converge, and SAC can faster stabilize at a higher reward than PPO, while its final reward value is lower than that of the proposed CODI approach. Furthermore, Fig. 4 (b) compares the total users' QoE obtained by three algorithms with respect to the amount of users. It can be observed that the proposed CODI approach is obviously superior to PPO and SAC.
This is mainly because the proposed approach can achieve better sampling quality and possess better long-term dependence procession capability. In particular, the proposed CODI approach can generate higher quality decisions (i.e., each user's resolution ratio and diffusion step) by iteratively denoising multiple times. Since each denoising step gradually adjusts the model's output, the effect of uncertainty and noise can be reduced, while the conventional DRL algorithms (i.e., SAC and PPO) directly map the given environment to the output in one step, and thus leading to inevitable deviations.

\begin{figure}[t]%
	\centering
	\subfloat[\tiny]{
		\includegraphics[width=.486\linewidth]{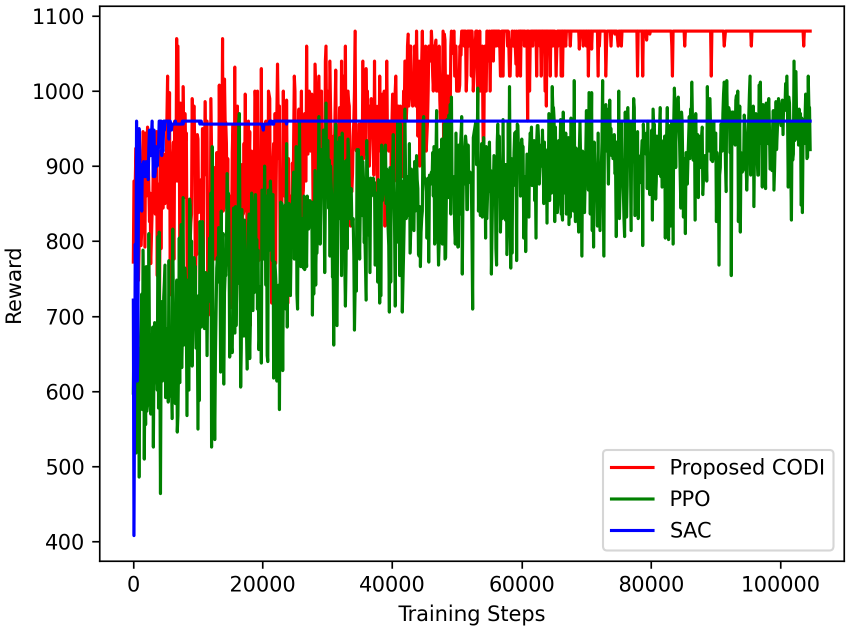} \label{subfig1}
		
	}\hfill \hspace{-1em}
	\subfloat[\tiny]{
		\includegraphics[width=.485\linewidth]{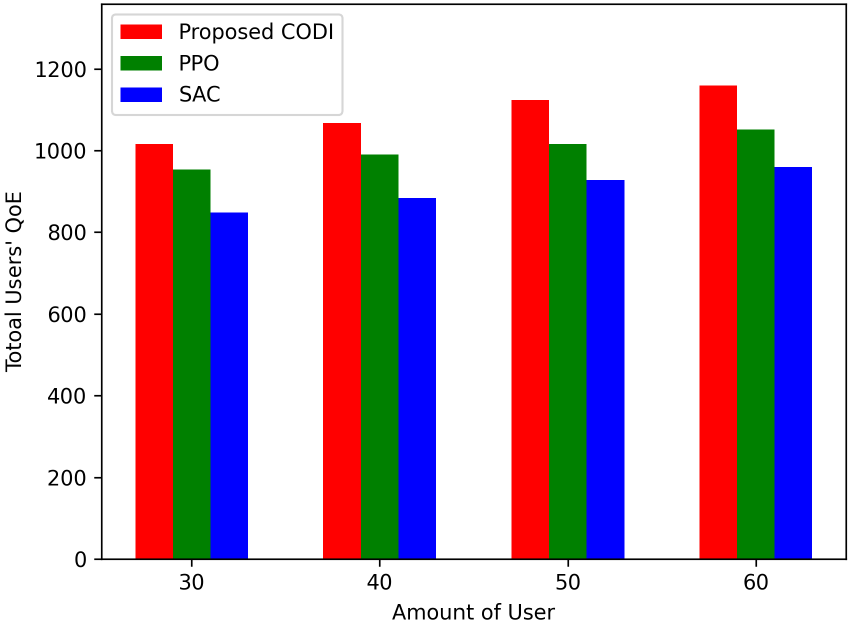} \label{subfig2}
	}
	\caption{(a) Comparison of the obtained reward w.r.t. number of training steps. (b) Comparison of the total users' QoE w.r.t. amount of user.}
\end{figure}

\section{Conclusions and Future Research Directions}
 In this article, we have presented a novel framework, namely mobile AIGC-driven HDT, for revolutionizing personalized healthcare. We propose the system architecture, and highlight the corresponding key design requirements and challenges for mobile AIGC-driven HDT. Moreover, we have illustrated two kinds of the potential applications and conducted a case study to demonstrate the effectiveness of mobile AIGC-driven HDT in a constructed testbed. However, there are still major open issues remaining to be studied in the future, as outlined below.

\begin{enumerate}
\item \emph{How to break down data silos}: The medical datasets used to pre-train mobile AIGC-driven HDT are collected from not only users themselves, but also many medical institutions owning larger amounts and more comprehensive historical data. However, the data and information in the current medical systems are fragmented and even severely segmented, resulting in data heterogeneity and bias. Specifically, medical data is relatively sensitive and to better protect patients' privacy, most medical institutions are unwilling to share their data. These indicate that it is imperative to develop a unified and secure data management framework, e.g., by integrating blockchain and federated learning, in medical systems to train a full-featured mobile AIGC-driven HDT.	
	
\item \emph{How to achieve cross-modal content generation}:
Cross-modal content generation for mobile AIGC-driven HDT is mainly challenged by the complexity of cross-modal matching and the lack of data standardization. Specifically, medical data of different modalities exhibit highly distinct features, making cross-modal matching tasks complicated. Besides, different medical institutions may use various equipment, each with its own standards, leading to the absence of standardized data formats or interfaces. To circumvent these difficulties, cross-domain collaborations with meta-learning and open set learning may be adopted to characterize the relationships among different data modalities, enhancing the model robustness and generalization.


\item \emph{How to balance tradeoff among high fidelity, accuracy and sustainability}: The updates of mobile AIGC-driven HDT involve repeated and asynchronous downloading and uploading high-dimensional (millions to billions) model parameters. This generates enormous data traffics and consumes a vast amount of energy in the communication and computation procedures. Thus, developing a robust, efficient, and sustainable training and deployment method for mobile AIGC-driven HDT is an urgent issue that needs to be addressed. Promising solutions may be green computing and communication, and model compression techniques, such as pruning, quantization and knowledge distillation.

\end{enumerate}

\bibliographystyle{IEEEtran}
\bibliography{IEEEabrv,AIGC_HDT}

\begin{thebibliography}{10}
\providecommand{\url}[1]{#1}
\csname url@samestyle\endcsname
\providecommand{\newblock}{\relax}
\providecommand{\bibinfo}[2]{#2}
\providecommand{\BIBentrySTDinterwordspacing}{\spaceskip=0pt\relax}
\providecommand{\BIBentryALTinterwordstretchfactor}{4}
\providecommand{\BIBentryALTinterwordspacing}{\spaceskip=\fontdimen2\font plus
\BIBentryALTinterwordstretchfactor\fontdimen3\font minus
  \fontdimen4\font\relax}
\providecommand{\BIBforeignlanguage}[2]{{%
\expandafter\ifx\csname l@#1\endcsname\relax
\typeout{** WARNING: IEEEtran.bst: No hyphenation pattern has been}%
\typeout{** loaded for the language `#1'. Using the pattern for}%
\typeout{** the default language instead.}%
\else
\language=\csname l@#1\endcsname
\fi
#2}}
\providecommand{\BIBdecl}{\relax}
\BIBdecl

\bibitem{1}
S.~D. Okegbile \emph{et~al.}, ``Human digital twin for personalized healthcare:
  Vision, architecture and future directions,'' \emph{IEEE Netw.}, pp. 1--7,
  2022.

\bibitem{10}
T.~Dong \emph{et~al.}, ``Privacy for free: How does dataset condensation help
  privacy?'' in \emph{Proc. ICML}, 2022, pp. 5378--5396.

\bibitem{4}
S.~Bond-Taylor \emph{et~al.}, ``Deep generative modelling: A comparative review
  of {VAEs, GANs}, normalizing flows, energy-based and autoregressive models,''
  \emph{IEEE Trans. Pattern Anal. Mach. Intell.}, vol.~44, no.~11, pp.
  7327--7347, 2022.

\bibitem{2}
J.~Chen \emph{et~al.}, ``Networking technologies for enabling human digital
  twin in personalized healthcare applications: A comprehensive survey,''
  \emph{arXiv preprint arXiv:2301.03930}, 2023.

\bibitem{16}
M.~AlAmir \emph{et~al.}, ``The role of generative adversarial network in
  medical image analysis: An in-depth survey,'' \emph{ACM Comput. Surv.},
  vol.~55, no.~5, pp. 1--36, 2022.

\bibitem{19}
A.~Kazerouni \emph{et~al.}, ``Diffusion models in medical imaging: A
  comprehensive survey,'' \emph{Med. Image Anal.}, p. 102846, 2023.

\bibitem{23}
P.~Prakash \emph{et~al.}, ``{RareBERT}: Transformer architecture for rare
  disease patient identification using administrative claims,'' in \emph{Proc.
  AAAI}, vol.~35, no.~1, 2021, pp. 453--460.

\bibitem{22}
T.~N. Jarada \emph{et~al.}, ``{SNF-CVAE}: computational method to predict
  drug--disease interactions using similarity network fusion and collective
  variational autoencoder,'' \emph{Knowl. Based Syst.}, vol. 212, p. 106585,
  2021.

\bibitem{8}
K.~S. Kim \emph{et~al.}, ``Ultrareliable and low-latency communication
  techniques for tactile internet services,'' \emph{Proc. IEEE}, vol. 107,
  no.~2, pp. 376--393, 2019.

\bibitem{18}
P.~Subramaniam \emph{et~al.}, ``Generating {3D TOF-MRA} volumes and
  segmentation labels using generative adversarial networks,'' \emph{Med. Image
  Anal.}, vol.~78, p. 102396, 2022.

\bibitem{24}
Y.~Song \emph{et~al.}, ``Solving inverse problems in medical imaging with
  score-based generative models,'' in \emph{Proc. ICLR}, 2022.

\bibitem{26}
T.~Uemura \emph{et~al.}, ``A generative flow-based model for volumetric data
  augmentation in {3D} deep learning for computed tomographic colonography,''
  \emph{Int. J. Comput. Assist. Radiol. Surg.}, vol.~16, pp. 81--89, 2021.

\bibitem{25}
J.~Wyatt \emph{et~al.}, ``{AnoDDPM}: Anomaly detection with denoising diffusion
  probabilistic models using simplex noise,'' in \emph{Proc. IEEE/CVF CVPR
  Workshops}, 2022, pp. 650--656.

\bibitem{27}
H.~Du \emph{et~al.}, ``{AI}-generated incentive mechanism and full-duplex
  semantic communications for information sharing,'' \emph{IEEE J. Sel. Areas
  Commun.}, 2023.

\bibitem{21}
T.~Li \emph{et~al.}, ``Applications of multi-agent reinforcement learning in
  future internet: A comprehensive survey,'' \emph{IEEE Commun. Surv. Tutor.},
  vol.~24, no.~2, pp. 1240--1279, 2022.

\end{thebibliography}

\vspace{-.5em}
\section*{Biography}

\small \textbf{Jiayuan Chen} is currently pursuing the Ph.D degree at the College of Computer Science and Technology, Nanjing University of Aeronautics and Astronautics, China. His research interests include edge computing and digital twin.

\small \textbf{Changyan Yi (Member, IEEE)} is a Professor with the College of Computer Science and Technology, Nanjing University of Aeronautics and Astronautics, China. His research interests include edge computing, industrial IoT, digital twin, 5G and beyond.

\small \textbf{Hongyang Du} is currently pursuing the Ph.D degree at the School of Computer Science and Engineering, Nanyang Technological University, Singapore. His research interests include semantic communications, generative AI and Metaverse.

\small \textbf{Dusit Niyato (Fellow, IEEE)} is a President's Chair Professor with the School of Computer Science and Engineering, Nanyang Technological University, Singapore. His research interests include edge intelligence, machine learning and incentive mechanism design.

\small \textbf{Jiawen Kang} is a professor with the School of Automation, Guangdong University of Technology, China. His research interests include Blockchain, IoT and Metaverse.

\small \textbf{Jun Cai (Senior Member, IEEE)} is a Professor and PERFORM Centre Research Chair with the Department of Electrical and Computer Engineering, Concordia University, Canada. His research interests include edge/fog computing and eHealth.

\small \textbf{Xuemin (Sherman) Shen (Fellow, IEEE)} is a University Professor with the Department of Electrical and Computer Engineering, University of Waterloo, Canada. His research focuses on network resource management, wireless network security, social networks and vehicular ad hoc networks. He is a Canadian Academy of Engineering Fellow, a Royal Society of Canada Fellow, and a Chinese Academy of Engineering Foreign Fellow.
\end{document}